\documentclass[prl,aps,showpacs,groupedaddress]{revtex4}
\usepackage{amssymb,amsmath,amsfonts,epsfig,graphicx,latexsym,bm,color}

\begin{document}
\title{Steering matter wave superradiance with an ultra-narrowband optical cavity}
\author{H. Ke{\ss}ler, J. Klinder, M. Wolke, and A. Hemmerich \footnote{e-mail: hemmerich@physnet.uni-hamburg.de} }
\affiliation{Institut f\"{u}r Laser-Physik, Universit\"{a}t Hamburg, Luruper Chaussee 149, 22761 Hamburg, Germany}
\date{\today}

\begin{abstract}A superfluid atomic gas is prepared inside an optical resonator with an ultra-narrow band width on the order of the single photon recoil energy. When a monochromatic off-resonant laser beam irradiates the atoms, above a critical intensity the cavity emits superradiant light pulses with a duration on the order of its photon storage time. The atoms are collectively scattered into coherent superpositions of discrete momentum states, which can be precisely controlled by adjusting the cavity resonance frequency. With appropriate pulse sequences the entire atomic sample can be collectively accelerated or decelerated by multiples of two recoil momenta. The instability boundary for the onset of matter wave superradiance is recorded and its main features are explained by a mean field model.
\end{abstract}


\maketitle
The coherent scattering of radiation by matter, commonly referred to as Rayleigh scattering, is an ubiquitous phenomenon in nature with basic consequences such as the blue color of the sky. If all scatterers are well localized within an optical wavelength of the incident radiation, their scattering contributions can sum up coherently, leading to a significant increase of the scattering cross section, a phenomenon closely related to the superradiance of collections of spontaneous emitters early discussed by Dicke \cite{Dic:54, Dic:64, Gro:82}. Even, if the sample by far exceeds sub-wavelength dimensions, scattered photons can imprint spatial correlations into the matter sample, which strongly enhance phase coherent scattering into certain directions, similarly as in Bragg scattering from material lattice structures. The use of ultracold gases has permitted to study superradiant Rayleigh scattering in the ultimate quantum mechanical limit when the atomic momentum is quantized in units of $\hbar k$ (with $k=2\pi/\lambda$, $\lambda \equiv$ optical wavelength of the irradiated light), a regime that has been termed matter wave superradiance \cite{Ino:99,Yos:04, Gil:07, Hil:08}. 

In the recent past remarkable progress has been made to tailor the light scattering properties of cold atomic matter ensembles in high finesse optical cavities \cite{Rit:13}. This has lead to promising new cavity-aided laser cooling methods \cite{Hor:97, Vul:00, Mau:04, Bei:05, Mor:07}, the observation of collective atomic recoil lasing \cite{Kru:03, Sla:07} and cavity enhanced Rayleigh scattering \cite{Bux:11} in ring cavities, or to the realization of atom-cavity systems showing extreme non-linear collective behavior, like optomechanical hysteresis and bistability \cite{Bre:08, Pur:10} or self-organization instabilities \cite{Dom:02, Cha:03, Bla:03, Bau:10, Arn:12, Sch:14}. In these experiments broad band cavities were used with linewidths well above the recoil frequency $\omega_{\mathrm{rec}} \equiv \hbar k^2/2m$ ($m\,=$ atomic mass) corresponding to the kinetic energy $E_{\mathrm{rec}} \equiv \hbar\,\omega_{\mathrm{rec}}$ gained by a resting atom after absorbing a single photon. 

\begin{figure}
\includegraphics[scale=0.5, angle=0, origin=c]{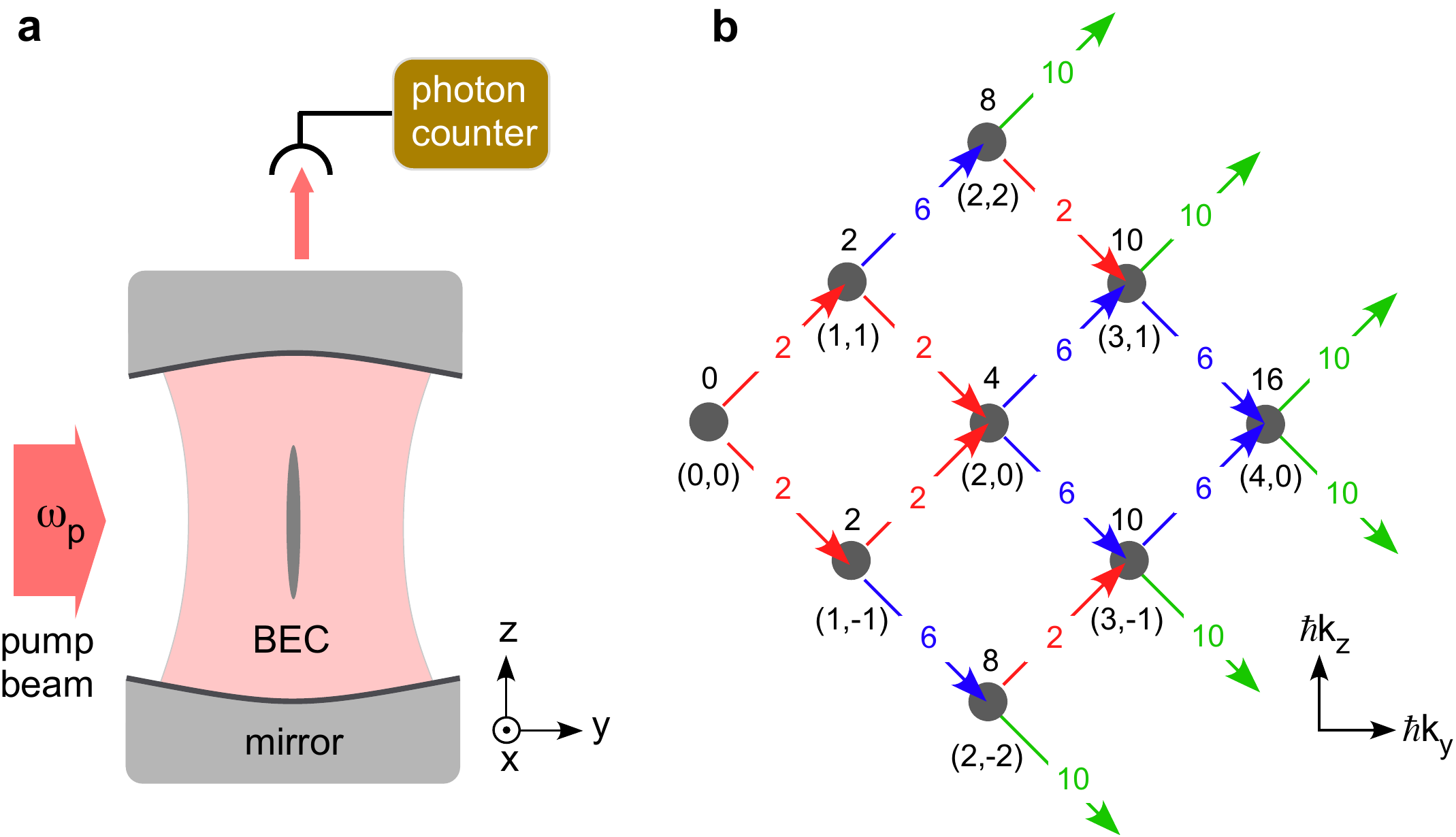}
\caption{\label{Fig.1} (color online). (a) A pump beam with frequency $\omega_p$ impinges upon a BEC inside a high finesse resonator. (b) Relevant momentum states coupled to the BEC by scattering photons from the pump beam. The black tuples $(n,m)$ below the filled grey disks denote the momenta of the respective momentum class along the $y$ and $z$ directions in units of $\hbar k$. The single numbers above the disks indicate the kinetic energy of the respective momentum class in units of the recoil energy. The colored numbers upon the arrows indicate the kinetic energy transfer associated with the respective scattering process.}
\label{fig:setup}
\end{figure}

In this work we investigate matter wave superradiance of a Bose-Einstein condensate (BEC) of rubidium atoms in the presence of a narrow band standing wave cavity, which combines sub-recoil energy resolution with a Purcell factor far above unity \cite{Pur:46, Tan:11}, such that the electromagnetic vacuum is significantly modified (see Fig.~\ref{fig:setup}a). Such cavities have been recently shown to open up new regimes of cavity cooling and cavity optomechanics \cite{Wol:12, Kes:14}. Here, we show that the use of such cavities in a Rayleigh scattering scenario, with a traveling pump wave irradiating the atoms perpendicularly with respect to the cavity axis, allows us to precisely address selected scattering channels, and thus to synthesize complex but yet well controlled, spatially periodic excited matter states preserving the full coherence of the initial condensate. We map out and explain the instability boundary for the onset of matter wave superradiance and discuss observations of suppression of superradiance, associated with destructive interference of different scattering channels. Remarkably, the single sided pumping of the atoms prevents the build-up of a stationary intra-cavity field even for negative detuning of the pump frequency with respect to the cavity resonance, in contrast to the observation of the Hepp-Lieb-Dicke phase transition \cite{Hep:73} for standing wave pumping \cite{Bau:10}. We use the cavity-aided control of Rayleigh scattering to demonstrate an efficient deceleration scheme for atoms, which could be also applied to other kinds of polarizable particles such as cold molecules.

In our experiment a cigar-shaped BEC of $N_a \approx 10^5$ $\mathrm{^{87}Rb}$-atoms is held in a magnetic trap with trap frequencies $\Omega_{\mathrm{x,y,z}} / 2\pi = (215.6 \times 202.2 \times 25.2)\,$Hz, irradiated by a pump beam propagating perpendicularly to the long axis of the condensate (see Fig.~\ref{fig:setup}(a)). The BEC with Thomas-Fermi radii $(3.1, 3.3, 26.8)\,\mu$m is prepared in the upper hyperfine component of the ground state $|{F=2,m_F=2}\rangle$. The single frequency ($\lambda = 803\,$nm) pump beam with a radius $w_p = 80 \, \mu$m is far detuned from the relevant atomic resonances (the atomic D$_{1,2}$ lines at $795\,$nm and $780\,$nm), such that its interaction with the atoms is dispersive with negligible spontaneous emission. The new element in our work is a high finesse narrowband optical cavity surrounding the BEC according to Fig.~\ref{fig:setup}(a). The field decay rate of $\kappa = 2\pi \times 4.5\,$kHz is smaller than $2\,\omega_{\mathrm{rec}} = 2\pi \times 7.1\,$kHz, which corresponds to the kinetic energy $2\,E_{\mathrm{rec}}$ transferred to a resting atom by scattering a pump photon into the cavity. The cavity axis is well aligned with the weakly confined $z$-axis of the condensate such that perfect spatial matching of the longitudinal TEM$_{00}$ cavity mode with the atomic sample is obtained. The high finesse of the TEM$_{00}$-mode ($\mathcal{F} = 3.44 \times 10^5$) together with its narrow beam waist ($w_0 \approx 31.2\, \mu$m) yield a Purcell factor $\eta_\mathrm{c} \approx 44$, i.e., scattering into the TEM$_{00}$ mode is enhanced by a factor 44 with respect to scattering into all other modes of the radiation field \cite{Pur:46, Tan:11}. Due to their preparation in the $|{F=2,m_F=2}\rangle$ hyperfine component of the ground state and the details of the D$_{1,2}$ lines, the maximal coupling to the atoms arises for left circularly polarized intra-cavity photons. For a uniform atomic sample, the TEM$_{00}$ resonance frequency for left circularly polarized light is dispersively shifted by an amount $\delta_{-} = \frac{1}{2} N_a \, \Delta_{-}$ with an experimentally determined light shift per photon $\Delta_{-} = \,2\pi \times 0.5\,$Hz (see Appendix). Hence, with $N_a = 10^5$ atoms $\delta_{-} = 2\pi\times 25$~kHz, which amounts to $5.6\,\kappa$, i.e., the cavity operates in the regime of strong cooperative coupling.

\begin{figure}
\includegraphics[scale=0.6, angle=0, origin=c]{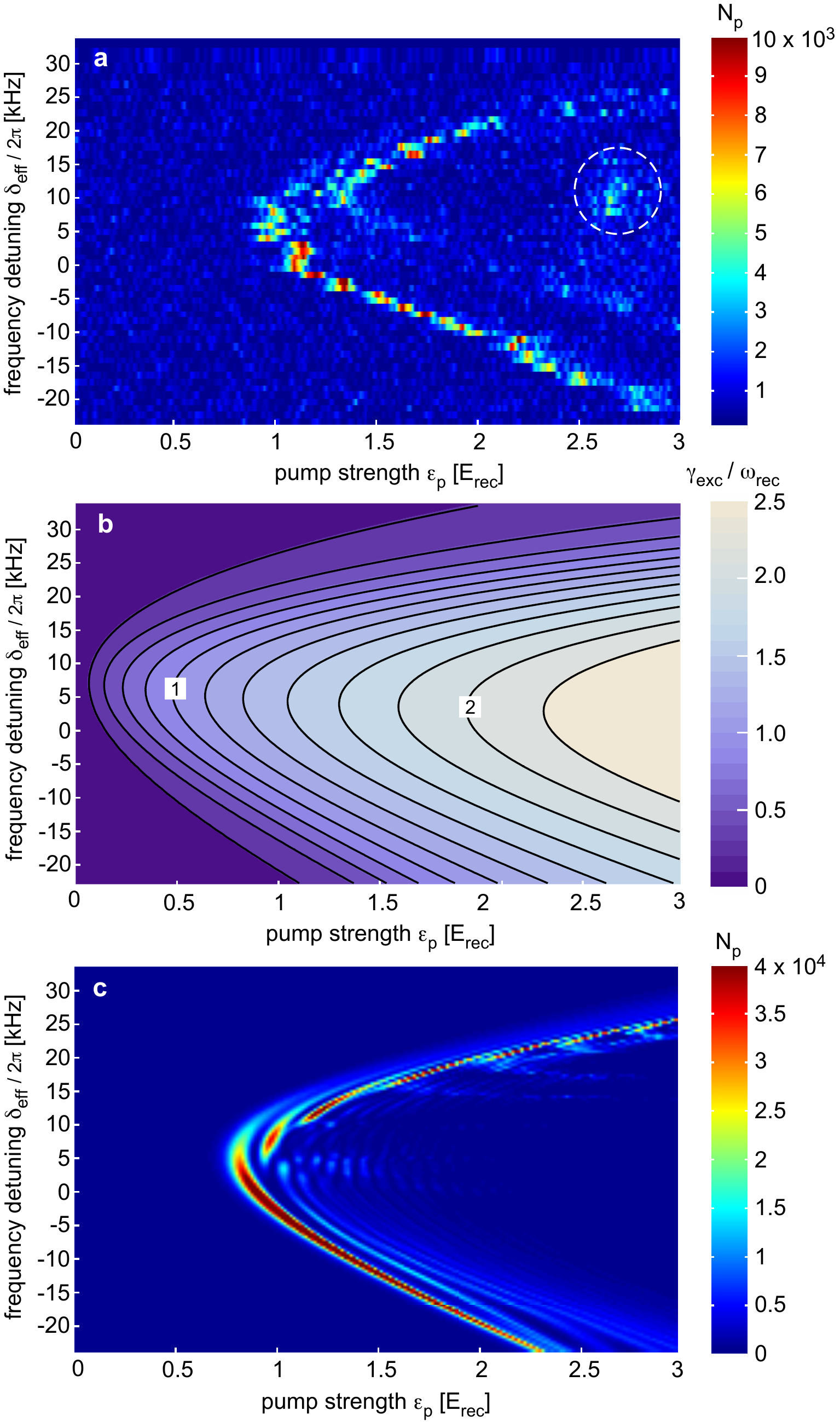}
\caption{\label{Fig.2} (color online). (a) The intra-cavity photon number $N_p$ is plotted versus the effective detuning $\delta_{\mathrm{eff}}/2\pi$ and the strength of the pump beam $\varepsilon_p/E_{\mathrm{rec}}$. The pump strength $\varepsilon_p$ is linearly increased from zero to 3 in 2 ms. As explained in the text, the signal within the dashed white circle is due to trap dynamics. (b) The excitation rate $\gamma_{\mathrm{exc}}$ characterizing the exponential instability is plotted versus $\delta_{\mathrm{eff}}/2\pi$ and $\varepsilon_p/E_{\mathrm{rec}}$. (c) Mean field simulation of the intra-cavity photon number for the pump strength ramp applied in (a).}
\label{fig:onset}
\end{figure}

As a consequence of the sub-recoil bandwidth, cavity assisted scattering can only occur in a narrow resonance window such that only very few selected motional states are coupled. This is sketched in Fig.~\ref{fig:setup}(b) for the simplified case when the transient formation of an intra-cavity optical lattice and the external trap are neglected and hence the atoms are considered as freely moving. Scattering of a pump photon by a BEC atom into the cavity corresponds to a transition from the $(0,0)\,\hbar k$ to the $(1,\pm 1)\,\hbar k$-momentum states. Energy conservation requires $\omega_p - \omega_{\mathrm{scat}} = 2\,\omega_{\mathrm{rec}}$ with $\omega_p$ and $\omega_{\mathrm{scat}}$ denoting the pump frequency and the frequency of the scattered photon, respectively. The scattering process is best supported by the cavity, if $\omega_{\mathrm{scat}}$ coincides with the effective cavity resonance frequency $\omega_{c,\mathrm{eff}} \equiv \omega_c - \delta_{-}$ with $\omega_c$ denoting the resonance frequency of the empty cavity, i.e., the effective detuning $\delta_{\mathrm{eff}} \equiv \omega_p - \omega_{\mathrm{c,eff}}$ should satisfy $\delta_{\mathrm{eff}} = 2\,\omega_{\mathrm{rec}}$. The same detuning allows to resonantly scatter a second photon bringing the atom to the $(2,0)\,\hbar k$ state. Further scattering, which would transfer the entire atomic sample via the $(3,\pm 1)\,\hbar k$ states to the $(4,0)\,\hbar k$ state, is not supported by the cavity unless $\delta_{\mathrm{eff}}$ is modified to account for the significantly larger energy costs of 6 recoil energies per atom. Due to the back-action of scattered photons upon the atomic sample, the scattering mechanism is expected to acquire collective character leading to the emission of a superradiant light pulse along the cavity axis: If the initial sample, a BEC in the $(0,0)\,\hbar k$ state, was perfectly homogeneous, scattering would be prevented by destructive interference from contributions from different locations within the BEC. Hence, quantum or thermal fluctuations are required to start the scattering process. Once a few photons are scattered into the cavity, the atoms transferred into the $(1,\pm 1) \hbar k$ momentum states form a standing matter wave commensurate with the weak optical standing wave potential produced in the cavity. The matter wave grating acts as a Bragg grating, which enhances the scattering efficiency such that the optical standing wave and the corresponding matter grating grow in an exponential process reaching maximal values, when most atoms populate $(1,\pm 1) \hbar k$. Their further transfer to $(2,0) \hbar k$ suppresses superradiance again, since in this state no density grating along the cavity axis is formed. 

\begin{figure}
\includegraphics[scale=0.4, angle=0, origin=c]{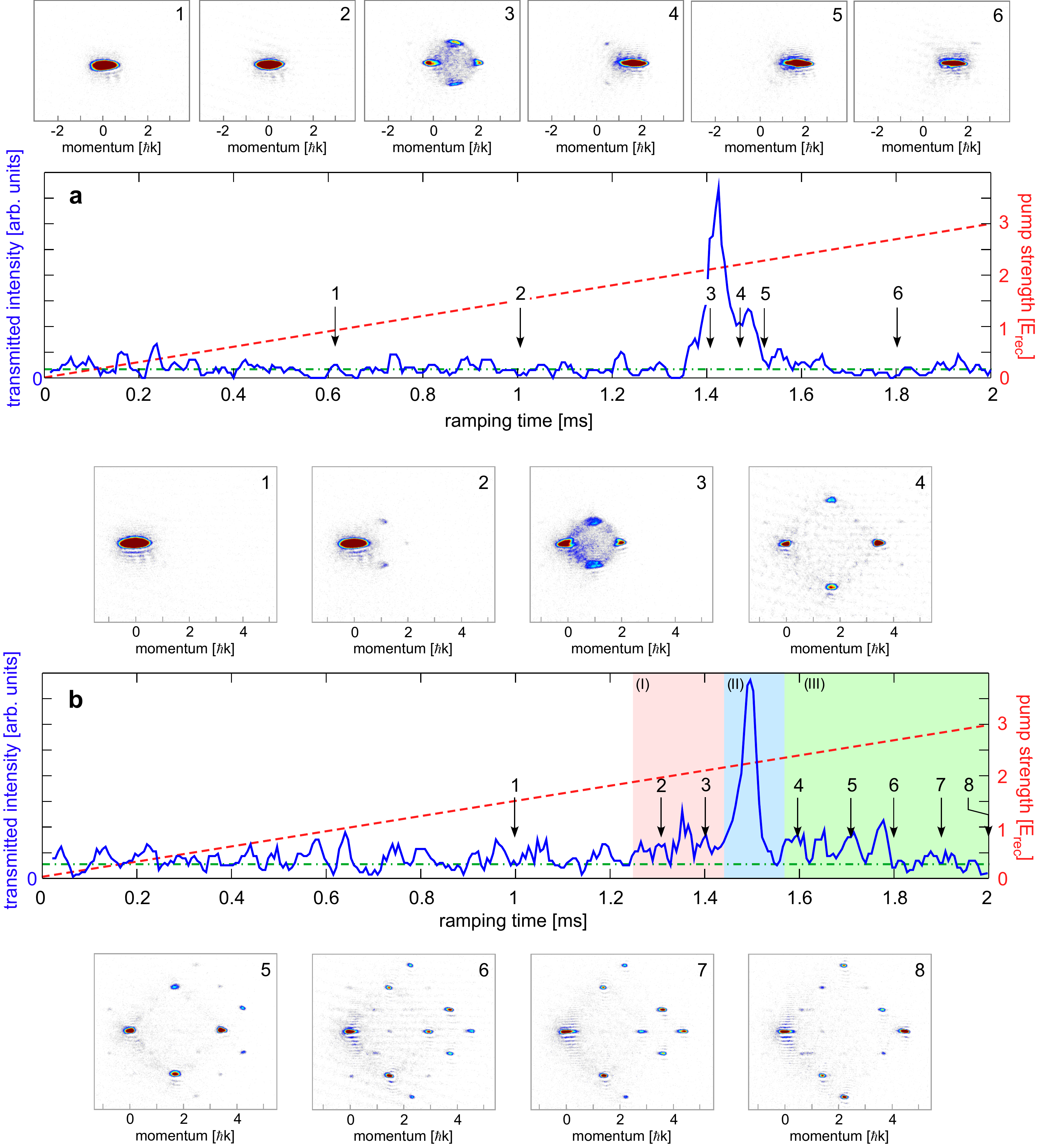}
\caption{\label{Fig.3} (color online). The blue solid traces show the intensity leaking out of the cavity, while the pump strength is ramped (as in Fig.~\ref{fig:onset}(a)) from zero to $3\,E_{\mathrm{rec}}$ in 2 ms (as indicated by the red dashed traces) with negative detuning $\delta_{\mathrm{eff}}/2\pi = -12\,$kHz in(a) and positive detuning $\delta_{\mathrm{eff}}/2\pi = 23\,$kHz in (b), respectively. The insets numbered 1 to 6 in (a) and 1 to 8 in (b) show (single shot) momentum spectra taken at times marked by the black arrows. The green dashed dotted lines indicate the noise floor of the light detection. For momentum spectra taken at late times in the pump strength ramp, the higher order momentum components are decelerated at the trap edge, which explains the compression of these spectra along the horizontal axis.}
\label{fig:transfer}
\end{figure}

The onset of superradiant scattering above a characteristic intensity of the pump beam is observed in our experiment. In Fig.~\ref{fig:onset}(a) we plot the intra-cavity power versus the frequency and the strength of the pump beam. The pump frequency is parametrized in terms of $\delta_{\mathrm{eff}}$ with $\delta_{\mathrm{eff}} = 0$ indicating the position of the cavity resonance in presence of the BEC. The pump strength is specified in terms of the peak light shift $\varepsilon_p$ caused by the pump beam in units of the recoil energy, which is spectroscopically measured (see Appendix). To derive the plot in Fig.~\ref{fig:onset}(a), the pump strength is increased linearly in time during 2 ms from 0 to 3 while the intra-cavity photon number is recorded by counting the photons leaking out through one of the mirrors. The observations show that at characteristic intensities, depending on $\delta_{\mathrm{eff}}$, short superradiant light pulses are emitted. Their locations in the $(\varepsilon_p, \delta_{\mathrm{eff}})$-plane take the approximate form of two nested parabolas rotated clockwise by $90^{\circ}$, which correspond to the instability boundaries for scattering associated with 2 and 6 $E_{\mathrm{rec}}$ kinetic energy transfer sketched by the red and blue arrows in Fig.~\ref{fig:setup}(b). The exact locations of these boundaries depend on the time used to ramp up the pump power. In our experiment, we are limited to a few milliseconds by the trap oscillation time $T_y = 2\pi /\Omega_{\mathrm{y}} \approx 5\,$ms for the $y$-direction. For ramping times approaching $T_y$ the scattered atoms are decelerated at the trap edge and hence are tuned back into resonance. This effect is responsible for the revival of intra-cavity intensity within the dashed white circle in Fig.~\ref{fig:onset}(a). The minimal pump strength required for superradiant scattering is found for $\delta_{\mathrm{eff}} \approx 2\,\omega_{\mathrm{rec}} = 2\pi \times 7.1\,$kHz, which corresponds to the expected resonance condition for the $(0,0)\,\hbar k \rightarrow (1,\pm 1)\,\hbar k$-transitions. 

The scattering instability can be understood by a simplified model only accounting for the $(0,0)\,\hbar k$ and $(\pm 1,\pm 1)\,\hbar k$ modes and neglecting depletion of the condensate (see Appendix). This model possesses a steady state solution with zero intra-cavity intensity and all atoms in the BEC at $(0,0)\,\hbar k$. This solution is unstable in the entire $(\varepsilon_p,\delta_{\mathrm{eff}})$ plane with an exponential excitation rate $\gamma_{\mathrm{exc}}(\varepsilon_p,\delta_{\mathrm{eff}})$ plotted in Fig.~\ref{fig:onset}(b). The graph shows that $\gamma_{\mathrm{exc}}$ is everywhere positive approaching zero on the $\delta_{\mathrm{eff}}$-axis. The contours indicate trajectories of constant $\gamma_{\mathrm{exc}}$ specified in units of $\omega_{\mathrm{rec}}$. These trajectories reflect the form of the instability boundary found in the experiment (Fig.~\ref{fig:onset}(a)). In particular, the value of $\delta_{\mathrm{eff}} = 2\,\omega_{\mathrm{rec}}$, which minimizes $\varepsilon_p$ along these trajectories agrees well with the observations. Note also the slight asymmetry with respect to the $\delta_{\mathrm{eff}} = 2\,\omega_{\mathrm{rec}}$ line, also observed in the experiment, which arises from contributions of the $(-1,\pm 1)\,\hbar k$ modes populated via re-scattering of cavity photons into the pump mode (see Appendix). A full mean field calculation of the intra-cavity photon number for the experimentally implemented 2~ms ramp of $\varepsilon_p$ is shown in Fig.~\ref{fig:onset}(c). The gross structure of the experimental data is nicely reproduced. 

\begin{figure}
\includegraphics[scale=0.4, angle=0, origin=c]{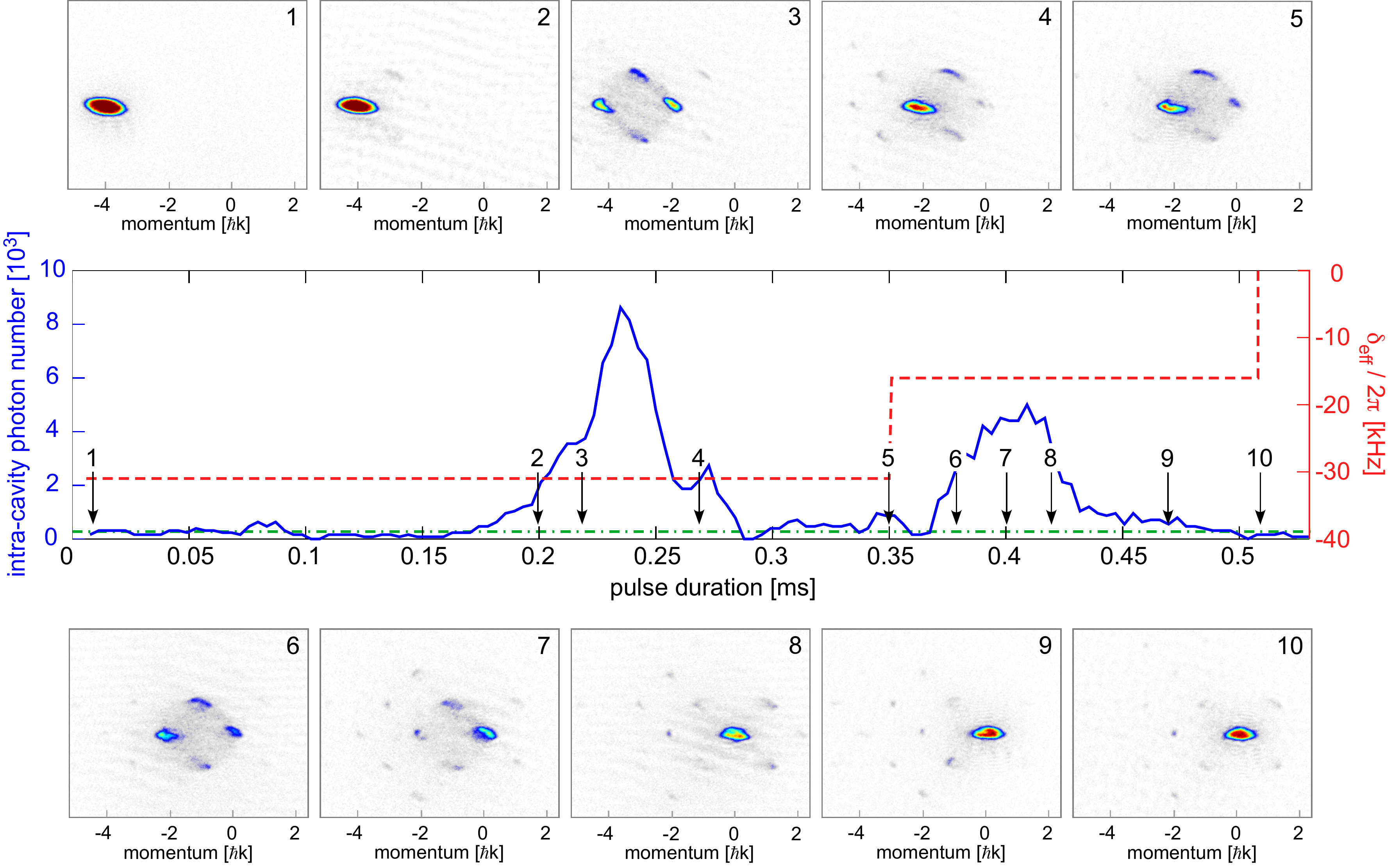}
\caption{\label{Fig.4} (color online). The blue solid traces shows the intra-cavity intensity for a sequence of two pump pulses with  $350\,\mu$s and  $160\,\mu$s durations and detunings $\delta_{\mathrm{eff}}/2\pi = -31\,$kHz and $\delta_{\mathrm{eff}}/2\pi = -16\,$kHz as indicated by the red dashed line. The pump strength was $\varepsilon_p = 2.5 E_{\mathrm{rec}}$ and $\varepsilon_p = 2.9 E_{\mathrm{rec}}$, respectively. The insets show momentum spectra at different times indicated by black arrows.}
\label{fig:deceleration}
\end{figure}

In Fig.~\ref{fig:transfer} we analyze the evolution of the momentum spectra of the atom sample corresponding to horizontal sections in Fig.~\ref{fig:onset}(a) for negative (a) and positive (b) detunings $\delta_{\mathrm{eff}}/2\pi = -12\,$kHz and $\delta_{\mathrm{eff}}/2\pi = 23\,$kHz. In (a) the intra-cavity intensity displays a single sharp superradiant spike, during which the BEC is completely transferred to the $(2,0)\,\hbar k$ momentum state as is illustrated by the momentum spectra shown in the insets. Conservation of energy requires $\omega_{\mathrm{scat}} - \omega_p = 2\,\omega_{\mathrm{rec}}$. Hence, with $\delta_{\mathrm{eff}}/2\pi = -12\,$kHz the frequency of the scattered light $\omega_{\mathrm{scat}}$ deviates from the effective cavity resonance frequency $\omega_{c,\mathrm{eff}}$ by $-4.3 \, \kappa$, i.e., the scattering processes $(0,0)\, \hbar k \rightarrow (1,\pm 1)\, \hbar k \rightarrow (2,0)\, \hbar k$ (red arrows in Fig.~\ref{fig:setup}(b)) are detuned from resonance. Other processes, however, are far more off-resonant: for a subsequent transfer to $(4,0)\,\hbar k$, requiring $\omega_{\mathrm{scat}}-\omega_p = 6\,\omega_{\mathrm{rec}}$ (blue arrows in Fig.~\ref{fig:setup}(b)), the detuning is $\omega_{\mathrm{scat}}-\omega_{c,\mathrm{eff}} = -7.5 \, \kappa$. Hence, after complete transfer to $(2,0)\, \hbar k$, scattering is blocked although the pump strength is further increased. 

The situation essentially changes, in the blue detuned case (b). The values of $\omega_{\mathrm{scat}} - \omega_{c,\mathrm{eff}}$ for processes transferring 2, 6, or 10 $E_{\mathrm{rec}}$ to the atoms (cf. red, blue, and green arrows in Fig.~\ref{fig:setup}(b)) are $(3.6, 0.37, -2.8) \kappa$. Hence, the transitions $(0,0)\, \hbar k \rightarrow (1, \pm 1)\, \hbar k$ and $(1, \pm 1)\, \hbar k \rightarrow (2,0)\, \hbar k$ with $2 E_{\mathrm{rec}}$ energy transfer are significantly slower than the nearly resonant processes, connecting to the $(2,0)\, \hbar k$ and $(1,\pm 1)\, \hbar k$ modes by transferring $6 E_{\mathrm{rec}}$ to the atoms. Therefore, in the experiment, first an increase of the intra-cavity intensity is seen (region highlighted by red background in Fig.~\ref{fig:transfer}(b), indicated by (I)), which results from scattering atoms from the condensate mode $(0,0)\, \hbar k$ into the $(1, \pm 1)\, \hbar k$ and $(2, 0)\, \hbar k$ modes. Once these modes become populated, the much faster nearly resonant processes with $6 E_{\mathrm{rec}}$ energy transfer set in (region highlighted in blue, (II)). The resulting rapid depletion of the $(1,\pm 1)\, \hbar k$ modes degrades the matter grating required to sustain the collective character of the $(0,0)\, \hbar k \rightarrow (2,0)\, \hbar k$ transition, which is therefore suppressed such that significant population can remain in the condensate mode as is seen in the inset (4) of Fig.~\ref{fig:transfer}(b). This suppression of superradiant scattering is related to the mechanism of subradiance in the Dicke model of spontaneous emitters \cite{Gro:82, DeV:96, Col:09}. Finally, in the region highlighted in green and labeled (III), the comparatively weaker processes, associated with $10 E_{\mathrm{rec}}$ energy transfer, populate an additional shell of momentum states, resulting in the small intra-cavity intensity level past $\approx 1.6\,$ms. The transfer of 14 $E_{\mathrm{rec}}$ would correspond to $\omega_{\mathrm{scat}} - \omega_{c,\mathrm{eff}} = -6 \kappa$, with the consequence that such processes are practically suppressed. As the insets in Fig.~\ref{fig:transfer}(b) show, only momentum states $(n, m)\,\hbar k$ become populated, which require at most $10\,E_{\mathrm{rec}}$ kinetic energy transfer per scattering event, i.e., $0 \leq n \leq 6, -3 \leq m \leq 3$ with the constraint $n+|m| \leq 6$.

Our findings demonstrate that the modification of the electromagnetic vacuum with a narrow band optical cavity provides control of the available momentum channels of superradiant Rayleigh scattering. This may, for example, be used to decelerate an initially moving BEC by a series of pump pulses with appropriately adjusted frequencies and intensities. In Fig.~\ref{fig:deceleration} a BEC, initially  prepared in the $(-4,0)\, \hbar k$ momentum state, is decelerated by two successive pump pulses: a first $350\,\mu$s long pulse with $\delta_{\mathrm{eff}}/2\pi = -31\,$kHz, $\varepsilon_p = 2.5\,E_{\mathrm{rec}}$ followed by a second pulse with $\delta_{\mathrm{eff}}/2\pi = -16\,$kHz, $\varepsilon_p = 2.9\,E_{\mathrm{rec}}$ and $160\,\mu$s duration. The insets depicting momentum spectra at different times during the pulse sequence show that the initial BEC after 0.5 ms is transferred to zero momentum. Note that due to collisions a significant amount of atoms is dispersed among a continuum of scattering states leading to the diffuse grey background. By adding additional pulses this scheme may be readily extended for deceleration of much faster particle samples. Other pulse sequences may be designed to implement efficient matter wave beam splitters or multiple path matter wave interferometers. An interesting future perspective of our work is the search for quantum entanglement between light and matter observables \cite{Nie:12}.

\begin{acknowledgments}
This work was partially supported by DFG-SFB 925 and DFG-GrK1355. We are grateful to Michael Thorwart, Reza Bakhtiari, Duncan O'Dell and Helmut Ritsch for useful discussions. We also thank Claus Zimmermann for his constructive critical reading of the manuscript.
\end{acknowledgments}

\section{Appendix}

\textbf{Parameters of Bose-Einstein condensate}.
A cigar-shaped Bose-Einstein condensate (BEC) with Thomas-Fermi radii $(3.1, 3.3, 26.8)\,\mu$m and $N_a \approx 10^5$ $\mathrm{^{87}Rb}$-atoms, prepared in the upper hyperfine component of the ground state $|F=2,m_F=2\rangle$, is confined by three centimeter-sized solenoids \cite{Han:06, Kli:10} arranged in a quadrupole Ioffe configuration \cite{Ess:98}, thus providing a magnetic trap with a nonzero bias field parallel to the $z$-axis with trap frequencies $\omega / 2\pi = (215.6 \times 202.2 \times 25.2)\,$Hz.
\\ \\
\textbf{Preparation of Bose-Einstein condensate with selected center-of-mass momentum.}
We start with a resting BEC magnetically trapped in the center of the cavity mode. With the help of an auxiliary coil the trap minimum and hence the BEC is adiabatically shifted along the y-axis by an adjustable amount $\Delta y$. Subsequently, the trap is rapidly shifted back to its original position, such that the BEC is now deflected from the trap center. Because the trap is harmonic across the range $\Delta y$, the only consequence is a harmonic oscillation of the BEC with the frequency $\omega_y$ such that after a waiting time $\tau = \frac{2\pi}{4\,\omega_y} = 1.24\,$ms we end up with an accelerated BEC at the trap center. By appropriate choice of $\Delta y$ we can precisely tune the center-of-mass momentum to $4\hbar k$.
\\ \\
\textbf{Cavity parameters}.
The high finesse of the standing wave cavity ($\mathcal{F} = 3.44 \times 10^5$) together with the narrow beam waist ($w_0 \approx 31.2\, \mu$m) yield a Purcell factor $\eta_\mathrm{c} \equiv \frac{24\,\mathcal{F}}{\pi\,k^2 w_0^2} \approx 44$ ($k \equiv 2\pi/\lambda$, and $\lambda=\,$ wavelength of the pump light). The cavity is oriented parallelly to the $z$-axis, such that the BEC is well matched to the mode volume of its TEM$_{00}$-modes. For a uniform atomic sample the resonance frequency for right ($+$) and left ($-$) circular photons is shifted due to the dispersion of a single atom by an amount $\Delta_{\pm} / 2$ with $\Delta_{\pm} = \frac{1}{2} \eta_\mathrm{c} \kappa \,\Gamma \left( \frac{f_{1,\pm}}{\delta_1} + \frac{f_{2,\pm}}{\delta_2}\, \right)$ and $\delta_{1,2}$ denoting the pump frequency detunings with respect to the relevant atomic D$_{1,2}$ lines at $780.2\,$nm and $795\,$nm \cite{Tan:11}. The quantities $\kappa = 2\pi \times 4.5\,$kHz and $\Gamma = 2 \pi \times 6$~MHz are the intra-cavity field decay rate and the decay rate of the $5\mathrm{P}$ state of $\mathrm{^{87}Rb}$, respectively. The prefactors $f_{1,\pm}$ and $f_{2,\pm}$ account for the effective line strengths of the D$_{1}$- and D$_{2}$-line components connecting to the $|F=2, m_F=2 \rangle$ ground state. The values of these factors are $(f_{1,-}, f_{2,-}) = (\frac{2}{3},\frac{1}{3})$ and $(f_{1,+}, f_{2,+}) = (0, 1)$. 

The quoted expressions for $\Delta_{\pm}$ use the rotating wave approximation and assume that the contributions from different transitions involved may be added. Finite size effects of the atomic sample and deviations of the intra-cavity field geometry from a plane wave are neglected. A more realistic value, used in our work, is obtained experimentally: The dispersive resonance shift for $N_a$ atoms $\frac{1}{2} N_a \Delta_{-}$ for left polarized light is measured by coupling a weak left polarized probe beam through one of the cavity mirrors to the TEM$_{00}$-mode. Its frequency is tuned across the resonance with and without atoms. At sufficiently low power levels of the probe the resonance is not affected by spatial structuring of the atoms due to backaction of the cavity field and hence merely results from the dispersion of the homogeneous sample. Accounting for the particle number $N_a$, know from absorption imaging, we find $\Delta_{-} \approx - 2\pi \times 0.5$~Hz corresponding to $\Delta_{-}/ \kappa \approx - 1.1 \times 10^{-4}$. Hence, with $N_a \approx 4 \times 10^{4}$ atoms the regime of strong cooperative coupling ($N\Delta_{-} > 4\kappa$) is entered.
\\ \\
\textbf{Pump beam parameters}. The pump beam with $w_p = 80\,\mu$m radius irradiates the BEC along the $y$-axis, i.e., perpendicularly with respect to its weakly confined $z$-axis. Its linear polarization is oriented parallelly to the $x$-axis and it operates at a wavelength $\lambda = 803\,$nm and therefore is detuned by $8\,$nm to the red side of the D$_{1}$-transition of $\mathrm{^{87}Rb}$. The pump strength is specified in terms of the magnitude of the peak light shift $\varepsilon_p \geq 0$ induced by the pump beam in units of the recoil energy. In order to calibrate the pump strength, the pump beam is retro-reflected in order to form a standing wave potential, the BEC is adiabatically loaded into this potential and the excitation spectrum is recorded and compared to a numerical band calculation to determine the antinode light shift. The pump strength parameter $\varepsilon_p$ is then defined as 1/4 times the measured antinode light shift. 

Our experiments require to tune the pump frequency with sub-kilohertz resolution across the resonance frequency of the TEM$_{00}$-mode interacting with the BEC. This is accomplished as follows (see also Ref.~\cite{Kes:14}): A reference laser operating at 803 nm is locked on resonance with a TEM$_{11}$-mode, which provides a cloverleaf-shaped  transverse profile. This mode exhibits a nodal line at the cavity axis such that the interaction with the BEC, which is positioned well in the center of the TEM$_{00}$-mode, is suppressed with respect to the TEM$_{00}$-mode by a geometrical factor $9 \times 10^{-5}$. Adjusting right circular polarization for the reference beam and hence $\sigma^+$-coupling to the BEC yields another suppression factor $\approx 0.43$. The pump laser, matched to couple the TEM$_{00}$-mode, is locked with an offset frequency of about $2.5\,$GHz to the reference laser. This offset is tunable over several MHz such that the vicinity of the resonance frequency of the TEM$_{00}$-mode can be accessed. 
\\ \\
\textbf{Detection of photons transmitted through the cavity}.
The light leaking out of the cavity is split into orthogonal circular polarization components and the photons of each component are counted with $56 \%$ quantum efficiency. The right circular photons, predominantly belong to the TEM$_{11}$-mode used to operate the stabilization of the pump beam frequency with respect to the cavity resonance Ref.~\cite{Kes:14}. Only a small fraction of these photons arises in the TEM$_{00}$-mode and results from the scattering of pump photons. For the experimental observations in Figs.~2, 3, and 4 of the main text the ratio between left and right circularly polarized photons found in the TEM$_{00}$-mode was 4. If the mirror transmission is known, the intra-cavity photon number can be determined. The mirror transmission was measured before its assembly in the cavity to be about 1 ppm. We suspect that this tiny transmission significantly varies across the mirror and hence may be easily smaller than 1 ppm at the position of the cavity mode. 
 \\ \\
\textbf{Mean field model}.
We consider a BEC of two-level atoms scattering light from an external traveling wave mode with the scalar electric field amplitude $\alpha_{0}(t) e^{iky}$ (pump mode) into a cavity mode with the scalar electric field $\alpha(t) \cos(kz)$. Neglecting collisional interaction the system is described by the set of mean field equations \cite{Rit:13}
\begin{eqnarray}
\label{eq:mean_field_model}
i\,\frac{\partial}{\partial t} \psi(y,z,t) &=& \left(-\frac{\hbar^2}{2m}\left[\frac{\partial^2}{\partial^2 y}+\frac{\partial^2}{\partial^2 z}\right] + \hbar \Delta_0 |\alpha(t) \cos(kz)+\alpha_{0}(t) e^{iky}|^2 \right) \psi(y,z,t) 
\\ \nonumber
\frac{\partial}{\partial t} \alpha(t) &=& \left(i \delta_c - i \Delta_0 \langle \cos^2(kz)\rangle_{\psi} - \kappa\right) \alpha(t) - i \Delta_0 \langle \cos(kz)\rangle_{\psi} \,\alpha_0(t) \, ,
\end{eqnarray}
with the matter wave function $\psi$ normalized to $N_a$ particles, and the electric fields normalized such that $|\alpha_{0}|^2$ and $|\alpha|^2$ denote the number of photons in the pump mode and the cavity mode, respectively. The light shift per intra-cavity photon is denoted by $\Delta_0$. $\langle \dots \rangle_{\psi}$ indicates integration over the BEC volume weighted with $|\psi|^2$. A plane wave expansion of $\psi(y,z,t)$ with respect to the relevant $(y,z)$-plane yields the corresponding scaled momentum space equations
\begin{eqnarray}
\label{eq:mean_field_equations}
\nonumber
i\,\frac{\partial}{\partial t} \phi_{n,m} &=& \omega_{\mathrm{rec}}\, \left(n^2+m^2 -2 |\beta|^2\, - \epsilon_p \right) \phi_{n,m} - \omega_{\mathrm{rec}} \,|\beta|^2 \, (\phi_{n,m-2}+\phi_{n,m+2}) 
\\ \nonumber
&+& i \,\omega_{\mathrm{rec}} \, \sqrt{\epsilon_p} \,\beta^{*}\, (\phi_{n-1,m-1} + \phi_{n-1,m+1}) - i\,\omega_{\mathrm{rec}}\,\sqrt{\epsilon_p} \, \beta \, (\phi_{n+1,m-1}+\phi_{n+1,m+1})
\\
\\  \nonumber
\frac{\partial}{\partial t} \beta &=& \left[i \, \left(\delta_{\mathrm{eff}} - \frac{1}{2} N_a \Delta_0 \sum_{n,m} \mathrm{Re}[\phi_{n,m}\phi_{n,m+2}^{*}] \right) - \kappa \right] \, \beta
\\ \nonumber
&-&\frac{1}{4} N_a \Delta_0 \,\sqrt{\epsilon_p} \, \sum_{n,m} \phi_{n,m}(\phi_{n+1,m-1}^{*}+\phi_{n+1,m+1}^{*}) \, ,
\end{eqnarray}
with $ \phi_{n,m}$ denoting the normalized ($\sum_{n,m}|\phi_{n,m}|^2 = 1$) amplitude of the momentum state $(n,m)\, \hbar k$. Upon the assumption of negative $\Delta_0$ the intra-cavity field $\beta$ is scaled such that $4 |\beta|^2 = - |\alpha|^2 \Delta_0 / \omega_{\mathrm{rec}}$ denotes the magnitude of the induced anti-node light-shift in units of the recoil energy. The pump strength parameter $\epsilon_p \equiv - |\alpha_0|^2 \Delta_0 / \omega_{\mathrm{rec}}$ is defined as the maximal light-shift induced by the pump beam in units of the recoil energy. The effective detuning is $\delta_{\mathrm{eff}}  \equiv \delta_c - \frac{1}{2} N_a \Delta_0$ with the detuning between the pump frequency and the empty cavity resonance $\delta_c$.  

A steady state solution of Eqs.~(\ref{eq:mean_field_equations}) is the zero solution $\beta = 0$ and $\phi_{n,m} = \delta_{n,0} \,\delta_{m,0}$. The stability properties of this solution may be studied by reducing Eqs.~(\ref{eq:mean_field_equations}) to the five matter modes $\phi_{0,0}$ and $\phi_{\pm 1,\pm 1}$. Switching to a basis such that the condensate has zero energy and neglecting its depletion, i.e., $\phi_{0,0} \approx 1$, one finds the system of linear equations
\begin{eqnarray}
\label{eq:3ModeModel}
i\,\frac{\partial}{\partial t} \left(\begin{array}{c} \beta \\ \beta^{*} \\ \phi_{+} \\ \phi_{+}^{*} \\  \phi_{-} \\ \phi_{-}^{*} \end{array} \right)
= 
\left(\begin{array}{cccccc}
- \delta_{\mathrm{eff}} - i \kappa & 0 & 0 &  i \, \lambda_{1}  & i \, \lambda_{1} & 0  \\
  0 & \delta_{\mathrm{eff}} - i \kappa & i \, \lambda_{1} & 0 & 0 &  i \, \lambda_{1} \\ 
  0 & i \lambda_{2} & 2\omega_{\mathrm{rec}} & 0 & 0 & 0 \\
  i \lambda_{2} & 0 & 0 & -2\omega_{\mathrm{rec}} & 0 & 0 \\
  -i \lambda_{2} & 0 & 0 & 0 & 2\omega_{\mathrm{rec}} & 0 \\
    0 & -i \lambda_{2} & 0 & 0 & 0 & -2\omega_{\mathrm{rec}}  \\
\end{array} \right) 
\left(\begin{array}{c} \beta \\ \beta^{*} \\ \phi_{+} \\ \phi_{+}^{*} \\  \phi_{-} \\ \phi_{-}^{*} \end{array} \right)
\end{eqnarray}
with the coupling parameters $\lambda_{1} \equiv -\frac{1}{4} \,N_a \Delta_0 \, \sqrt{2 \epsilon_p}$, $\lambda_{2} \equiv \omega_{\mathrm{rec}}\,\sqrt{2 \epsilon_p}$ and $\phi_{\pm} \equiv \frac{1}{\sqrt{2}}\left(\phi_{\pm 1,1} + \phi_{\pm 1,-1} \right)$, which formally resembles a Schr{\"o}dinger equation for a six-level system with a non-Hermitian Hamiltonian \cite{Ben:13} giving rise to imaginary Rabi-frequencies. If the imaginary part of one of the eigenvalues of the matrix on the right hand side of Eq.~(\ref{eq:3ModeModel}) is positive, an exponential instability arises and hence the system is rapidly driven away from the zero solution. 

In order to compare experimental observations with the model in Eq.~(\ref{eq:mean_field_equations}) and Eq.~(\ref{eq:3ModeModel}), the experimental parameters $\Delta_{\pm}, \varepsilon_p$ and the model parameters $\Delta_{0}, \epsilon_p$ must be connected accounting for the fact that in the model two-level atoms are assumed and the vectorial character of the electric field is neglected. In the experiment, the strongest coupling to the atoms arises for left circular light with respect to the natural quantization axis fixed by the magnetic off-set field along the $z$-axis. Hence, we identify $\Delta_{0} = \Delta_{-}$. Inside the cavity, the linear $\hat{x}$-polarization of the pump beam may be decomposed into equally strong left and right circular components with respect to the $z$-axis. Only the left circular component can scatter into the left circularly polarized cavity mode. Hence, the light shift $\varepsilon_p$ induced by the pump beam in the experiment is related to the number of pump photons $|\alpha_0|^2$ used in the model description by $\varepsilon_p = - 2 |\alpha_0|^2 (\Delta_{+} + \Delta_{-}) / \omega_{\mathrm{rec}}$ and thus $\varepsilon_p/\epsilon_p = (\Delta_{+} + \Delta_{-}) /\Delta_{0} = 1.44$. A more involved description, which is deferred to forthcoming work, should account for two orthogonal polarization modes of the cavity operating with different effective detunings. Hence, at the present stage, precise quantitative agreement between the model and the observations is not to be expected.

In Figure 2(b) of the main text the maximum of the imaginary parts of the six eigenvalues of the matrix on the right hand side of Eq.~(\ref{eq:3ModeModel}) is plotted versus $\delta_{\mathrm{eff}}$ and $\varepsilon_p$. It is interesting to note the asymmetry with respect to the $\delta_{\mathrm{eff}} = 2\,\omega_{\mathrm{rec}}$ line, which is also observed in the experiment. It arises due to the interaction between the matter modes $\phi_{+}$ and $\phi_{-}$ mediated via their interaction with the intra-cavity mode $\beta$. The matter mode $\phi_{-}$ cannot be directly populated by the pump beam via scattering by a condensate atom, because pump photons can only transfer momentum into the $y$ but not the $-y$-direction. Population of $\phi_{-}$ can only arise, if a photon scattered into the cavity is re-scattered back into the pump beam before it is lost through a cavity mirror. In fact, if these processes are neglected by setting $\phi_{-}$ to zero, an instability boundary is calculated, which is perfectly mirror symmetric with respect to the $\delta_{\mathrm{eff}} = 2\,\omega_{\mathrm{rec}}$ line.

Figure 2(c) in the main text was obtained by solving Eqs.~(\ref{eq:mean_field_equations}) for a linear ramp of $\varepsilon_p$ with 2 ms duration including all modes with $-4 \leq n,m \leq 4$. A small initial deviation from $\phi_{0,0} =1$ is required in order to leave the unstable zero solution. In the experiment, this deviation is naturally provided by thermal or quantum fluctuations. We assumed that the first excited modes $(\pm 1,\pm 1)\,\hbar k$ are populated according to a Boltzman factor with a temperature $T = 0.2 \,T_c$ ($T_c =$ critical temperature of the BEC). Hence, we set $\phi_{0,0} = \cos(\theta)$ and $\phi_{\pm 1, \pm 1} = \sin(\theta)/2$ with $\theta = \arctan(2 e^{-\hbar \omega_{\mathrm{rec}} / k_B T})$. We have checked that the gross structure of the solution does not depend on the exact choice of the temperature. The agreement with the observation in Figure 2(a) is qualitative. The calculated maximal photon numbers notably exceed the measured values, while the calculated superradiant light pulses are shorter than their observed counterparts. The former may reflect our unreliable knowledge of the outcoupling mirror transmission. The quantitative discrepancies may also result from simplifications made in our model. A more realistic calculation should include both available circularly polarized intra-cavity modes, which are coupled to the incident linearly polarized pump photons with different strength and effective resonance frequencies. Furthermore, the finite size of the pump beam should be accounted for, the radius of which is only a factor 2.5 larger than the radius of the BEC such that atoms at the edge of the BEC experience decreased values of $\varepsilon_p$. Finite size effects leading to a Mie-scattering scenario may also be significant \cite{Bac:12}. Another important issue in our experiment is collisional interaction, since due to the tight external trap the particle densities are well above $10^{14}\,$cm$^{-3}$. This leads to significant populations in a continuum of collision states, which decreases the number of atoms contributing to superradiant scattering.

\end{document}